\newif\ifconf

\newif\iffull
\fulltrue 

\ifconf
\documentclass[a4paper,UKenglish,cleveref, autoref, thm-restate]{lipics-v2021}
\fi

\iffull
\documentclass[11pt]{article}
\usepackage{fullpage,utopia}
\date{}
\fi

\usepackage{amsfonts,amssymb,amsthm,amsmath,comment}
\usepackage{hyperref}
\usepackage{cleveref}

\iffull
\newtheorem{theorem}{Theorem}

\newtheorem{lemma}[theorem]{Lemma}

\newtheorem{observation}[theorem]{Observation}
\newtheorem{corollary}[theorem]{Corollary}
\newtheorem{claim}[theorem]{Claim}

\theoremstyle{definition}

\newtheorem{definition}[theorem]{Definition}
\newtheorem{remark}[theorem]{Remark}

\newenvironment{claimproof}[1]{\par\noindent{\em Proof of Claim.}~#1}
\fi

\newcommand{\NP}{\mbox{\small\rm{NP}}}
\newcommand{\W}{\mbox{\small\rm{W}}}
\newcommand{\SAT}{\textsc{Sat}}
\newcommand{\twosat}{\text{$2$-\textsc{Sat}}}
\newcommand{\ksat}{\text{$k$-\textsc{Sat}}}
\newcommand{\usa}{\textsc{USA}}
\newcommand{\paf}{\textsc{PAF-Sat}}
\newcommand{\pafsat}[1]{{#1}-\textsc{PAF-Sat}}

\newcommand{\poleq}{\textsc{Poly-Eqs}}
\newcommand{\polyeq}[1]{{#1}\textsc{-Poly-Eqs}}
\newcommand{\Subsat}{\textsc{Sub-Sat}}
\newcommand{\subsat}[1]{{#1}\textsc{-Sub-Sat}}
\newcommand{\sat}[1]{{#1}\textsc{-Sat}}
\newcommand{\subssat}{\textsc{Sub-Sat}}

\newcommand{\maxsubsat}{\textsc{Max-2-Sub-Sat}}
\newcommand{\maxesubsat}{\textsc{Max-E2-Sub-Sat}}
\newcommand{\maxpafsat}{\textsc{Max-2-PAF-Sat}}
\newcommand{\maxepafsat}{\textsc{Max-E2-PAF-Sat}}

\newcommand{\F}{\mathbb{F}}

\DeclareMathOperator{\depth}{depth}
\DeclareMathOperator{\poly}{poly}
\DeclareMathOperator{\codim}{codim}

\DeclareMathOperator{\2cnf}{2\mbox{-}CNF}

\newcommand{\kcnf}{\text{$k$-CNF}}
\newcommand{\cnf}{\text{CNF}}
\newcommand{\horn}{\text{HORN}}

\renewcommand{\ge}{\geqslant}

\renewcommand{\le}{\leqslant}

\newcommand{\multclique}{\textsc{Multicolored-Clique}}
\newcommand{\oxr}{\mathsf{OXR}}

\def\out{\text{out}}
\def\inn{\text{in}}

\title{CNF Satisfiability in a Subspace and Related Problems}

\iffull

\author{V. Arvind\thanks{Institute of Mathematical Sciences (HBNI),
    Chennai, x1India. Email: {\tt arvind@imsc.res.in}} \and Venkatesan Guruswami\thanks{Computer Science Department, Carnegie
  Mellon University, Pittsburgh, USA. Email: {\tt venkatg@cs.cmu.edu}. Portions of this work were done during visits to the Institute of Mathematical Sciences, Chennai. Research supported in part by the US National Science Foundation grant CCF-1908125 and a Simons Investigator Award.}}
  \fi
\ifconf
\titlerunning{Satisfiability in a Subspace}

\author{V. Arvind}{Institute of Mathematical Sciences (HBNI), Chennai,
  India}{email:arvind@imsc.res.in}{}{}

\author{Venkatesan Guruswami}{Computer Science Department, Carnegie
  Mellon University, USA}{email: venkatg@cs.cmu.edu}{}{Portions of this work were done during visits to the Institute of Mathematical Sciences, Chennai. Research supported in part by the National Science Foundation grant CCF-1908125 and a Simons Investigator Award.}

\authorrunning{V. Arvind and V. Guruswami}

\ccsdesc[500]{Theory of computation~Parameterized complexity and exact algorithms}

\keywords{CNF Satisfiability, Exact exponential algorithms, Hardness results}

\category{} 

\relatedversion{} 

\fi

\begin{document}

\maketitle
\thispagestyle{empty}

\begin{abstract}
   We introduce the problem of finding a satisfying assignment to a
   CNF formula that must further belong to a prescribed input
   subspace. Equivalent formulations of the problem include finding a
   point outside a union of subspaces (the Union-of-Subspace Avoidance
   (USA) problem), and finding a common zero of a system of
   polynomials over $\F_2$ each of which is a product of affine forms.
    
   \smallskip We focus on the case of $\kcnf$ formulas (the
   $\subsat{k}$ problem). Clearly, $\subsat{k}$ is no easier than
   $k$-SAT, and might be harder. Indeed, via simple reductions we show
   that $\subsat{2}$ is \NP-hard, and $\W[1]$-hard when parameterized
   by the co-dimension of the subspace. We also prove that the
   optimization version Max-$\subsat{2}$ is \NP-hard to approximate
   better than the trivial $3/4$ ratio even on satisfiable instances.
    
\smallskip On the algorithmic front, we investigate fast exponential
algorithms which give non-trivial savings over brute-force
algorithms. We give a simple branching algorithm with runtime $(1.5)^r$
for $\subsat{2}$, where $r$ is the subspace dimension, as well as an
$O^*(1.4312)^n$ time algorithm where $n$ is the number of variables.
    
    \smallskip
    Turning to $\subsat{k}$ for $k \ge 3$, while known algorithms for
    solving a system of degree $k$ polynomial equations already imply
    a solution with runtime $\approx 2^{r(1-1/2k)}$, we explore a more
    combinatorial approach.  Based on an analysis of critical
    variables (a key notion underlying the randomized $k$-SAT
    algorithm of Paturi, Pudlak, and Zane), we give an algorithm with
    runtime $\approx {n\choose {\le t}} 2^{n-n/k}$ where $n$ is the
    number of variables and $t$ is the co-dimension of the
    subspace. This improves upon the runtime of the polynomial
    equations approach for small co-dimension. Our combinatorial approach also achieves polynomial space in contrast to the algebraic approach that uses exponential space.
    We also give a
    PPZ-style algorithm for $\subsat{k}$ with runtime $\approx
    2^{n-n/2k}$. This algorithm is in fact oblivious to the structure
    of the subspace, and extends when the subspace-membership
    constraint is replaced by any constraint for which partial
    satisfying assignments can be efficiently completed to a full
    satisfying assignment.  Finally, for systems of $O(n)$ polynomial
    equations in $n$ variables over $\F_2$, we give a fast exponential
    algorithm when each polynomial has bounded degree irreducible
    factors (but can otherwise have large degree) using a degree
    reduction trick.
    \end{abstract}

\newpage

\section{Introduction}\label{intro}

Given an $n$-variate Boolean formula $\Phi$ along with an affine
subspace $A\subseteq \F_2^n$ (given by a system of $\F_2$-linear
equations) as input, we explore the complexity of testing if $\Phi$
has a satisfying assignment in $A$. This is a natural twist on Boolean
constraint satisfaction problems that studies the effects of linear
algebra on Boolean logic.  Our focus shall be on the case when $\Phi$
is presented in Conjunctive Normal Formal (CNF). We refer to this
problem as \emph{satisfiability in a subspace} and denote it by
$\subssat$.  This framework can capture non-Boolean problems such as
Graph $K$-Colorability indicating the richness of combining the
problem of Boolean CNF-satisfiability with a linear-algebraic
constraint. We also note that in the area of practical SAT solvers
there is interest in CNF satisfiability conjuncted with XOR
constraints \cite{SoosM19,SoosGM20}.

Further, $\subssat$ has two other equivalent interesting
formulations. The first of these is \emph{union of subspace
  avoidance}, $\usa$ for short: Given affine subspaces
$A_1,A_2,\ldots,A_m\subseteq \F_2^n$ is there an $x\in\F_2^n$ that is
not in the union $\bigcup_{i=1}^m A_i$?  A different formulation is a
special case of finding a solution to a bunch of polynomial equations
$p_i=0$ over $\F_2^n$, namely when each $p_i$ is a product of affine
forms. We refer to this reformulation as $\paf$. We will describe
these (easy) equivalences in Section~\ref{sec:prelims}.

For most of the paper, we restrict attention to the case when $\Phi$
is a $k$-CNF formula (a CNF formula with clauses of width at most $k$)
for a fixed $k$, referred to as the $\subsat{k}$ problem. Clearly,
$\subsat{k}$ is a generalization of the well-studied \ksat\ ($k$-CNF
satisfiability). In terms of the two reformulations above, $\subsat{k}$
corresponds to the $\usa$ problem when the spaces $A_i$ have
co-dimension at most $k$, and for the $\paf$ problem, each polynomial
$p_i$ is the product of up to $k$ affine forms.

We present both hardness results and algorithms for $\subsat{k}$, described in Sections~\ref{intro-hard} and \ref{intro-algos} below respectively. Owing to the NP-hardness of the problems, the algorithmic focus is on exponential time algorithms that give non-trivial improvements over brute-force.

There are two possible angles from which to view the study of $\subsat{k}$. The first is as a problem intermediate between satisfiability of $k$-CNF formula and a system of degree $k$ polynomial equations. The second is as a specific instance of a constraint satisfaction problem (CSP) obtained by combining two fundamental types of constraints. There have been a few works~\cite{PW10,CS16} giving algorithms beating brute-force for some natural problems with mixed constraints, but we are still far from a general picture of how to obtain fast exponential algorithms for a combined template of constraints when each constraint type does admit such non-trivial algorithms. In this context, tackling the combination of $k$-CNF formulas and linear equations is a good starting point, and one that could hopefully spur a more systematic study in the future. There have been a few investigations~\cite{JLNZ17,LW18,CHL19,JLR21} into the fine-grained complexity of CSPs via the algebraic approach based on (partial) polymorphisms. This theory has developed the tools to compare the optimal exponents of different constraint types, identifying for instance the ``easiest" NP-hard CSP within some classes. However, with the exception of \cite{BG19}, polymorphisms have not been leveraged to design fast exponential algorithms with competitive exponents.

\subsection{Hardness results}\label{intro-hard}

Since $\subsat{k}$ is a generalization of \ksat, $\subsat{k}$
inherits all the intractability results of \ksat\ for $k \ge 3$. This
leaves the interesting case of $\subsat{2}$. This turns out to be much
harder than the polynomial time solvable $\sat{2}$. We establish the
following, showing not just hardness (even for FPT algorithms) of the
exact version, but also a tight inapproximability for the
approximation version (even on satisfiable instances). The proofs are
based on short, simple reductions, once an appropriate problem to
reduce from is chosen.\footnote{The \NP-hardness would also follow from Schaefer's
dichotomy theorem for Boolean CSP~\cite{Sc78}, though that is an
overkill hammer for this result.} The $\W[1]$-hardness answers a question posed in \cite{BGG-stoc19} on the fixed-parameter complexity of $\sat{2}$ with a global modular constraint, parameterized by the modulus.

\begin{theorem}
\label{thm:2sat-hardness-intro}
\begin{enumerate}
\item $\subsat{2}$ is $\NP$-hard. It is further $\W[1]$-hard when
  parameterized by the co-dimension of the affine space $A$ in which
  we seek a satisfying assignment.
\item Given a satisfiable instance of $\subsat{2}$, it is \NP-hard to
  find an assignment in the input space $A$ that satisfies more than
  $3/4+\epsilon$ of the 2SAT clauses, for any $\epsilon > 0$.
\end{enumerate}
\end{theorem}

\subsection{Algorithmic results}\label{intro-algos}

Analogous to seeking \ksat\ algorithms faster than brute-force, we
investigate fast exponential time algorithms for $\subsat{k}$ that
beat the naive brute-force $2^{\dim(A)}$ time algorithm, where
$A\subseteq \F_2^n$ is the subspace in which we seek a
solution. Algorithms for \ksat\ have received much attention and are
central to the burgeoning field of fast exponential-time
algorithms. The algorithmic theory is closely connected to fixed
parameter tractability and parameterized complexity \cite{FG,DF}. The
accompanying hardness theory \cite{IP01,IPZ01}, based on the
exponential-time hypothesis (ETH) and the strong exponential-time
hypothesis (SETH), is a sanity check to the quest for faster
algorithms for \ksat \ and other NP-complete problems.

There are several interesting \ksat\ algorithms with running time
$O^*(2^{n(1-\Theta(1/k))})$.\footnote{The notation $O^*(f(n))$ for
  runtime bounds suppresses polynomial factors.}  We only mention two
significant algorithms from among these: one by Paturi, Pudlak,
Zane~\cite{PPZ} and another due to Sch\"oning~\cite{Schoning99}. Both
algorithms are simple to describe with delightfully clever and elegant
analyses. The PPZ algorithm considers variables in a random order, and
gives each a random value unless its value is forced by a clause and
previously set values. It achieves a runtime of $O^*(2^{n(1-1/k)})$.
Sch\"{o}ning's algorithm starts with a random assignment and in each
step fixes an unsatisfied clause by flipping the value of a random one
of its variables. It achieves a runtime of $O^*((2-2/k)^n)$.

Given that $\subsat{k}$ generalizes \ksat, it is natural to seek
exponential algorithms with similar runtimes for $\subsat{k}$.  For
$\subssat$ with input space $A \subseteq \F_2^n$, the brute-force
algorithm in fact runs in time $O^*(2^{\dim(A)})$. A natural question
is whether we can get similar improvements in the exponent of the
$O^*(2^{\dim(A)})$ runtime.

An algorithm~\cite{LPTWY17} with running time about $O^*(2^{r(1-
  1/5k)})$ is known for checking satisfiability of a collection of
arbitrary degree $k$ polynomial equations in $r$ variables: Let
$P_i\in\F_2[x_1,x_2,\ldots,x_r]$, $1\le i\le m$, be polynomials over the
field $\F_2$. Following \cite{LPTWY17}, the $\poleq$ problem is
solving the system of polynomial equations $P_i=0, 1\le i\le m$ over
$\F_2$: to check if there exists a solution in $\F_2^r$ and compute
one if it exists. When $P_i$ are all of degree bounded by $k$ we
denote this special case by $\polyeq{k}$. The $\polyeq{k}$ problem
generalizes $\subsat{k}$ by the following easy transformation: Suppose
the subspace $A$ where we seek a satisfying assignment is $r$
dimensional. Then we can express the $i^{th}$ clause in the
$\subsat{k}$ instance as a disjunction of $k$ affine linear forms in
$r$ variables: $C_i=(\ell_{i,1} \vee \ell_{i,2} \vee\cdots\vee \ell_{i,k})$.  We
define the corresponding polynomial $P_i=\prod_{j=1}^k (\ell_{i,j}+1)$.
Now, the $\subsat{k}$ instance is satisfiable iff the $\polyeq{k}$
instance $P_i=0, 1\le i\le m$ has a solution in $\F_2^r$.

The algorithm \cite{LPTWY17} is a novel application of the
Razborov-Smolensky ``polynomial method,'' originally developed as a
lower bound technique, used to define low-degree probabilistic
polynomials for approximating the OR gate. The same idea allows for
replacing a system of polynomial equations by a single probabilistic
polynomial (without significant increase in degree), followed by a
partial table lookup search.  The article \cite{LPTWY17} presents more
general results applicable to all finite fields $\F_q$. Recently, in
\cite{D21}, the running time for the case of $\F_2$ has been improved
to $O^*(2^{r (1- 1/2k)})$ by a refinement of the search method in
\cite{LPTWY17}.


Since $\subsat{k}$ is a special case of solving a system of polynomial
equations over $\F_2$, it raises the natural question of improving the
running time further to match the $O^*(2^{r(1-1/k)})$ runtime of the
PPZ randomized algorithm for $\ksat$. We are only able to achieve
this speed-up in some special cases. However, on the positive side, our algorithms turn out to be \emph{ polynomial space bounded}, unlike the polynomial equations based method which requires exponential space \cite{LPTWY17,D21}.

\subsubsection{Algorithms for $\subsat{2}$} For $\subsat{2}$ a simple deterministic branch-and-bound
algorithm achieves a runtime of $O^*(3^{r/2})$ where $r$ is the dimension of the subspace $A$. We can improve
on this with a randomized branching strategy to a runtime of $O^*(1.5^r)$. This improves over the randomized $O^*(1.6181^r)$
algorithm given by the polynomial method~\cite{D21} for solving a system of quadratic equations over $\F_2$. There is also a simple deterministic branching algorithm with $O^*(((1+\sqrt{5})/2)^r)$ runtime for $\subsat{2}$. This is based on the same branching strategy for $\ksat$ \cite[Theorem, pp.~295]{MS85} with its runtime governed by the generalized Fibinacci numbers.

When $\dim(A)=n-t$, we can adapt the
  algorithm from \cite[Algorithm 4.1]{BGG-stoc19} (for solving 2-SAT with a single
  abelian group constraint) to obtain an $O^*({n \choose {\le t}})$ time
  algorithm.
  \footnote{For nonnegative integers $n,t$, the notation ${n \choose {\le t}}$ stands for $\sum_{i=0}^t {n \choose i}$.}  

  The result of Theorem~\ref{thm:2sat-hardness-intro} shows that this
  problem is not in FPT parameterized by the co-dimension $t$,
  answering a question posed in \cite{BGG-stoc19} on whether 2-SAT
  with a global abelian group constraint might be fixed-parameter
  tractable, parameterized by the group size. More generally, the
  work~\cite{BGG-stoc19} systematically studied the effect of a global
  modular constraint on the complexity of Boolean constraint
  satisfaction problems, exposing many interesting phenomena and
  connections.

Balancing the two runtimes of $O^*(1.5^r)$ and $O^*({n \choose n-r})$ algorithm when $r \ge n/2$ (the exponents of the two bounds become equal at $r=(1-\eta)n$ for $\eta \approx 0.115816$) yields a
$O^*(1.4312^n)$ time randomized algorithm for $\subsat{2}$ on $n$
variables. The following records these results.
  
\begin{theorem}\label{thm:ksat-algo1}
 There is a randomized $O^*(1.5^r)$ algorithm for $\subsat{2}$ where
 $r$ is the dimension of the input space, as well a deterministic $O^*({n \choose {\le t}})$ time algorithm where $t$ is the co-dimension. 
 Together, these imply a randomized
 $O^*(1.4312^n)$ time algorithm as a function of the number $n$ of variables.
\end{theorem}

\subsubsection{Algorithms for $\subsat{k}$}

We explore combinatorial algorithms for $\subsat{k}$ based on the
notion of \emph{critical variables} (which was introduced in
\cite{PPZ} and plays an important role in their satisfiability
algorithm). Let $\Phi$ be a satisfiable $\cnf$ formula in $n$
variables $x_i,i\in[n]$, and let $\bar{a}\in\F_2^n$ be a satisfying
assignment.

\begin{definition}{\rm \cite{PPZ}}\label{cr-def}
We say $x_i$ is a \emph{critical variable} for $\bar{a}$ with respect
to $\Phi$ if the assignment $\bar{a}+e_i$ falsifies $\Phi$, where
$e_i$ is the $i^{th}$ elementary vector with $1$ in the $i^{th}$
coordinate and zero elsewhere (so $\bar{a}+e_i$ is just $\bar{a}$ with
$x_i$ flipped). If the formula $\Phi$ is clear from context, we simply
say that $x_i$ is a critical variable for assignment $\bar{a}$.
\end{definition}
    
The key idea in our combinatorial algorithms is \emph{plucking} of
non-critical variables based on the following simple observation: if
$\Phi$ is an $n$-variate $\cnf$ formula and $\bar{a}$ is a satisfying
assignment such that variable $x_i$ is non-critical for it, then the
formula $\Phi'$ obtained by plucking $x_i$ (i.e., dropping all
occurrences of $x_i$ and its complement from $\Phi$) remains
satisfiable with $\bar{a}'\in\F_2^{n-1}$ as a satisfying assignment,
where $\bar{a}'$ is obtained from $\bar{a}$ by dropping the $i^{th}$
coordinate.

The important property of $\Phi'$ is that given any satisfying
assignment for $\Phi'$ we can set $x_i$ to either $0$ or $1$ to
recover a satisfying assignment for $\Phi$. This facilitates searching
for a satisfying assignment in an affine space $A$: if the plucked
variable $x_i$ occurs in a linear constraint defining $A$ then we can
drop that linear constraint while seeking a satisfying assignment for
$\Phi'$, because that linear constraint can always be satisfied by
choosing the right value of $x_i$ which still remains overall a
satisfying assignment for $\Phi$. Based on this idea we obtain the
following algorithms for $\subsat{k}$:
\begin{itemize}
\item The first result here is a randomized $O^*({n\choose
  {t}}2^{n-n/k})$ time algorithm for $\subsat{k}$ where $t=\codim(A)$.
  This algorithm is essentially governed by the running time of the
  PPZ satisfiability algorithm \cite{PPZ} combined with an iterative
  ``search and pluck'' operation to remove $t$ non-critical variables
  from the $t$ linear equations defining $A$. This running time is
  superior to the $O^*(2^{r-r/2k})$ time randomized algorithm based on
  solving polynomial equations for small values of $t=o(n)$.

\item The second result is a general randomized
  $O^*(2^{n-n/2k+n/2k^2})$ time algorithm for $\subsat{k}$, nearly
  matching the $\approx 2^{r-r/2k}$ run time of the polynomial
  equations algorithm \cite{D21,LPTWY17} for $r$ close to $n$.  It
  again uses the PPZ satisfiability algorithm as a subroutine combined
  with simple applications of the plucking step: if the number of
  critical variables is fewer than $n/2$, it randomly guesses and
  plucks non-critical variables. This algorithm does not need to look
  at the linear equations defining $A$. In fact, it works for any
  Boolean constraint $C(x_1,x_2,\ldots,x_n)$ (replacing membership in
  the affine space $A$) with a polynomial-time algorithm that takes a
  partial assignment and extends it to an assignment that satisfies
  $C$. For example, $C$ can be a $\horn$ or dual $\horn$ formula.
  
\item It is pleasing to note that we can apply the idea of plucking
  non-critical variables to $\subsat{2}$ and obtain an $O^*({n\choose
    {\le t}})$ deterministic algorithm (cf.~\cite{BGG-stoc19}), where
  $t=\codim(A)$. Exploiting the structure of $\2cnf$ formulas, we can
  find the non-critical variables efficiently.
\end{itemize}

\begin{theorem}\label{thm:ksat-algo1-intro}
The $\subsat{k}$ problem admits two randomized algorithms, one running
in time $O^*(2^{n-n/2k+n/2k^2})$, and another running in
$O^*({n\choose {t}}2^{n-n/k})$ when the input subspace has
co-dimension $t \le n/2$.\footnote{Of course, there is also a trivial
  $O^*(2^{n-t})$ time brute force algorithm.} Both algorithms use space bounded by a polynomial in $n$.
\end{theorem}

\begin{remark}
Satisfiability algorithms based on the switching lemma (which converts \kcnf\ to decision trees of moderate term size and number of terms) are known in the literature (e.g., see \cite{IMP12}). We can easily adapt this algorithm to solve
$\subsat{k}$, because once we have a decision tree for the
underlying $\kcnf$ formula, for the $\subsat{k}$ instance each path of the decision tree will give rise to a system of linear equations over $\F_2$. For each path, therefore, we can even count the number of satisfying assignments. Counting over all the paths of the decision tree gives the total number of satisfying
assignments for the $\subsat{k}$ instance in randomized time $O^*(2^{n(1-1/c\cdot k)})$ for some suitable large constant $c>0$. Furthermore, the algorithm is also polynomial space-bounded. In terms of running time, however, it is 
a much weaker bound in comparison to \cite{LPTWY17} or even the algorithms of Theorem~\ref{thm:ksat-algo1-intro}. In this context, we note that for $\# \sat{k}$ there is a \emph{deterministic} $O^*(2^{n(1-1/c\cdot k)})$ time algorithm based on the polynomial method (albeit using exponential space)~\cite{ChW21}. We do not know of any such deterministic algorithm for counting satisfying assignments to $\subsat{k}$. 
\end{remark}

 Finally, motivated by the (unbounded $\cnf$) $\Subsat$ problem, we
 revisit the general problem solving a system of polynomial equations
 $p_i=0,1\le i\le m$ over $\F_2$, where $m=O(n)$, where each $p_i$ is
 given by an arithmetic circuit of $\poly(n)$ degree. In the case when
 each $p_i$ has small degree irreducible factors, we get a
 $2^{r(1-\alpha)}$ time randomized algorithm, where $\alpha$ depends
 on the number of equations $m$ and the degree bound on the
 irreducible factors (Theorem~\ref{thm:schuler-type}).

\subsection{Equivalent and related problems to $\subssat$}\label{sec:prelims}

Recall the $\usa$ problem: Given a collection of affine subspaces
$A_1,A_2,\ldots,A_m\subseteq \F_2^n$ (where each $A_i$ is given by a
bunch of affine linear equations over $\F_2$) the problem is to
determine if there is a point $x\in \F_2^n\setminus \bigcup_{i=1}^m
A_i$. 

Clearly, the complement $\F_2^n\setminus \bigcup_{i=1}^m A_i$ is
expressible as an $\textrm{AND}$ of $\textrm{OR}$s of affine linear
forms $\oplus_{i\in S}x_i+b$, $b\in\{0,1\}$. Thus, $\usa$ is clearly
reducible to $\subssat$. The converse reduction is also easy: given a
CNF formula $\Phi$ and an affine subspace $A\subseteq\F_2^n$ we first
convert it to an $\textrm{AND}$ of $\textrm{OR}$s of affine linear
forms. An assignment $x\in A$ satisfies $\Phi$ if and only if it
satisfies $C_1\wedge C_2\wedge\cdots\wedge C_m$, where each clause
$C_i$ is an $\textrm{OR}$ of affine linear forms. The set $A_i$ of
satisfying assignments of the complement $\overline{C_i}$ is an affine
subspace of $\F_2^n$, and $\Phi$ is satisfiable by $x\in A$ if and
only if $x\in \F_2^n\setminus \bigcup_{i=1}^m A_i$.

For the equivalence to $\paf$, suppose $\Phi = C_1\wedge
C_2\wedge\cdots\wedge C_m$, where each clause $C_i$ is an
$\textrm{OR}$ of affine linear forms $C_i = \vee_{j=1}^t L_{ij}$.  As
already discussed in Section~\ref{intro-algos}, the assignment
$x\in\F_2^n$ satisfies $C_i$ if and only if it satisfies the
polynomial equation $\prod_{j=1}^m(L_{ij}+1) = 0$. Thus, the
satisfiability of $\Phi$ is reducible to a system of $m$ polynomial
equations $p_i=0$, where each $p_i$ is a product of affine linear
forms. The converse reduction is also easy which we omit.

\paragraph*{Organization of the paper.}  We present the
results in a different order than in the introduction. In
Section~\ref{sec-algos1} we first present the algorithms for $\subsat{k}$ and then for $\subsat{2}$. In Section~\ref{sec-hard} we present our hardness
results for $\subsat{2}$. Finally, in Section~\ref{sec-algos2} we
present the algorithm for $\poleq$ for $O(n)$ equations $p_i=0$, where
each $p_i$ has unrestricted degree but constant-degree irrreducible
factors. 

\section{Algorithmic results for $\subsat{k}$}\label{sec-algos1}

As mentioned in the introduction, the $\subsat{k}$ problem seems
intermediate in difficulty, between $\ksat$ and the problem
$\polyeq{k}$ of solving a system of degree-$k$ polynomial equations
over $\F_2$. The latter problem has an $O^*(2^{r(1-1/2k)})$ time
algorithm \cite{LPTWY17,BKW19,D21}, which yields an
$O^*(2^{r(1-1/2k)})$ time algorithm for $\subsat{k}$, where
$r=\dim(A)$.

Ideally, we would like an algorithm for $\subsat{k}$ with run time
$O^*(2^{r(1-1/k)})$, with savings in the exponent similar to that of
the PPZ algorithm \cite{PPZ} for \ksat.

We present some algorithms in this direction: For $\subsat{2}$ there
is a simple $O^*(1.5^r)$ time randomized algorithm which improves on
the $O^*(2^{r(1-1/2k)})$ bound for $k=2$.  For a special case of
$\subsat{k}$, when $r=\dim(A)$ is close to the number of variables
$n$, we are able to adapt the PPZ algorithm to essentially get an
$O^*(2^{r(1-1/2k)})$ time algorithm. Writing $t=n-r=\codim(A)$, we can
even obtain an $O^*({n\choose \le t}\cdot 2^{n(1-1/k)}$ time algorithm for
the problem, also based on the PPZ satisfiability algorithm, which
yields the desired $1/k$ savings in the exponent for small $t$.

\subsection{An $O^*({n\choose t}\cdot 2^{n(1-1/k)})$ time randomized algorithm: co-dimension $t$ case}\label{sec-algo1}

As outlined in Section~\ref{intro-algos}, the algorithm will use the
PPZ satisfiability algorithm \cite{PPZ} as a subroutine, combined with
variable plucking steps to solve $\subsat{k}$ in randomized time
$O^*({n\choose t}\cdot 2^{n(1-1/k)})$, when $\codim(A)=t$.  In
particular, for $\codim(A)=o(n)$ the algorithm has run time
$O^*(2^{n(1-1/k+o(1))})$.

The variable plucking is based on analyzing the critical variables for
a solution $\bar{a}\in\F_2^n$ of a given $\subsat{k}$ instance
$(\Phi,A)$, depending on whether or not they occur in the linear
equations defining $A$.

For an instance $(\Phi,A)$ we partition the variables into two sets
\[
\{x_i\mid i\in[n]\} = V_\inn\sqcup V_\out,
\]
where $V_\inn$ is the subset of variables that have nonzero
coefficient in at least one of the $t$ linear equations defining $A$,
and $V_\out$ is the remaining set of variables. By abuse of notation,
we will also treat $V_\inn\sqcup V_\out$ as a partition of the index
set $[n]$. We consider the following two cases.

\medskip\noindent \textbf{Case 1.} Suppose $(\Phi,A)$ has the property
that \emph{for every solution} $\bar{a}\in\F_2^n$ each variable in
$V_\inn$ is critical for $\bar{a}$ w.r.t $\Phi$.  There is no variable
plucking required in this case. It only involves the application of
the PPZ satisfiability algorithm on $\Phi$ and checking that the
assignment found belongs to $A$. We need the following lemma which is
analogous to \cite[Lemma 4]{PPZ}.  The proof of the lemma is by an
induction argument like in \cite{PPZ}.

\begin{lemma}\label{clm:ppz-influence}
Let $S$ be a nonempty subset of $\F_2^n$.  For each $\bar{a}\in S$,
let $I_{\out}(\bar{a})=\{i\in V_{\out}\mid \bar{a}+e_i\notin S\}$,
where $e_i$ is the $i^{th}$ elementary vector. Then we have
\begin{equation}
\label{eq:ppz-isoperimetry}
\sum_{\bar{a}\in S}2^{|I_{\out}(\bar{a})|-|V_{\out}|}\ge 1.
\end{equation}
\end{lemma}

\iffull
\begin{proof}
  If $|V_{\out}| = 0$, then $I_{\out}(\bar{a}) = \emptyset$ for every
$\bar{a} \in S$, and the left hand side of \eqref{eq:ppz-isoperimetry}
equals $|S|$ which is at least $1$.

So assume $|V_{\out}| \ge 1$ and without loss of generality that $1\in
V_{\out}$.  Let $S_0 = \{ \bar{a} \in S \mid a_1=0\}$ and $S_1 = \{
\bar{a} \in S \mid a_1=1\}$, and also denote $V'_{\out} = V_{\out}
\setminus \{1\}$.

First consider the case when both $S_0$ and $S_1$ are nonempty. For
$\bar{a} \in S_0$, define $I^{(0)}_{\out}(\bar{a}) = \{ i \in
V'_{\out} \mid \bar{a} +e_j \notin S_0\}$ and likewise for $\bar{a}
\in S_1$, define $I^{(1)}_{\out}(\bar{a}) = \{ i \in V'_{\out} \mid
\bar{a} +e_j \notin S_1\}$.  By induction hypothesis, applied w.r.t
$V'_{\out}$, and pairs $S_0$ and $I^{(0)}_{\out}(\bar{a})$, as well as
$S_1$ and $I^{(1)}_{\out}(\bar{a})$, we know that
\begin{equation}
    \label{eq:ind-hyp}
 \sum_{\bar{a}\in S_0}2^{|I^{(0)}_{\out}(\bar{a})|-|V'_{\out}|}\ge 1 \quad \text{and} \quad 
 \sum_{\bar{a}\in S_1}2^{|I^{(1)}_{\out}(\bar{a})|-|V'_{\out}|}\ge 1 \ .
\end{equation}

Now if index $j \in I^{(0)}_{\out}(\bar{a})$ for some $\bar{a} \in
S_0\subset S$, then $\bar{a} + e_j \notin S_0$ and as the first
coordinate of $\bar{a}+e_j$ is also $0$, we have $\bar{a} + e_j \notin
S$, and thus $j \in I_{\out}(\bar{a})$. Thus $|I_{\out}(\bar{a})| \ge
|I^{(0)}_{\out}(\bar{a})|$ for all $\bar{a} \in S_0$. Likewise,
$|I_{\out}(\bar{a})| \ge |I^{(1)}_{\out}(\bar{a})|$ for all $\bar{a}
\in S_1$. Since $|V'_{\out}| = |V_{\out}| -1$, using these in
\eqref{eq:ind-hyp}, we conclude \eqref{eq:ppz-isoperimetry} in this
case, as desired.

Next, suppose $S=S_0$ and $S_1 = \emptyset$ (the case when $S_0
=\emptyset$ is handled the same way). In this case, for every $\bar{a}
\in S$, $1 \in I_{\out}(\bar{a})$, as $S_1 = \emptyset$ and thus
flipping the first bit will always lead to a vector outside $S$. Thus
$|I_{\out}(\bar{a})| = |I^{(0)}_{\out}(\bar{a})|+1$. Using this
together with $|V'_{\out}| = |V_{\out}|-1$ in the first inequality of
\eqref{eq:ind-hyp}, we conclude \eqref{eq:ppz-isoperimetry} in this
case as well.
\end{proof}
\fi

Now, let $\bar{a}\in\F_2^n$ be some solution of the $\subsat{k}$
instance $(\Phi,A)$. Then, by the assumption of Case 1 and the
preceding discussion $\bar{a}$ has $|V_\inn| + |I_\out(\bar{a})|$
critical variables w.r.t $\Phi$.

Following the analysis in \cite{PPZ}, if we now run one iteration of
the PPZ algorithm on the instance $\Phi$, the probability that
$\bar{a}$ is output is at least
\[
\frac{1}{n^2}\cdot 2^{-n+(|V_\inn| + |I_\out(\bar{a})|)/k}.
\]

Let $S\subset \F_2^n$ denote the subset of solutions to the instance
$(\Phi,A)$.  Summing up over all $\bar{a}\in S$, the probability that
some solution $\bar{a}$ is output is given by
\begin{eqnarray*}
  \sum_{\bar{a}\in S} \frac{1}{n^2}\cdot 2^{-n+(|V_\inn|/k +
    |I_\out(\bar{a})|/k)}& = & \frac{1}{n^2}2^{-n+n/k}\cdot \sum_{\bar{a}\in S} 2^{(-|V_\out|/k + |I_\out(\bar{a})|)/k}\\
  &\ge& \frac{1}{n^2}2^{-n+n/k}\cdot \sum_{\bar{a}\in S} 2^{(-|V_\out| + |I_\out(\bar{a})|)} \ge  \frac{1}{n^2}2^{-n+n/k} \ ,
\end{eqnarray*}
where the last step uses Lemma~\ref{clm:ppz-influence}. This finishes
the analysis of Case 1.

\begin{remark}
Notice in the probability analysis that $S$ is the set of solutions to
$(\Phi,A)$ and not all solutions to $\Phi$. The crucial property that
for every $\bar{a}\in S$, each variable in $V_\inn$ is critical w.r.t
$\Phi$ yields that there are $|V_\inn| + |I_\out(\bar{a})|$ critical
variables for $\bar{a}$ w.r.t $\Phi$. Intuitively, as the variables in
$V_\out$ do not occur in the linear equations, the PPZ algorithm when
run on $\Phi$ will be able to deterministically set, on average,
$|I_\out(\bar{a})|/k$ many of the critical variables in $V_\out$
without any interaction with the linear equations defining $A$.
\end{remark}

\medskip\noindent \textbf{Case 2.}
We now consider the case when not all variables in $V_\inn$ are
critical to all solutions to $(\Phi,A)$.  We will show that there is a
subset of at most $t$ variables in $V_\inn$ that can be plucked from
$\Phi$ and reduce the transformed instance to Case 1. We will argue
that the algorithm can do an exhaustive search for this subset of
$V_\inn$ of size at most $t$.  

\begin{lemma}\label{case2to1}
In the $\subsat{k}$ instance $(\Phi,A)$, let $Bx=b$ be the system of
$t$ linear equations defining $A$. Suppose variable $x_1$ occurs in
the first equation $\sum_{j=1}^n B_{1j}x_j=b_1$ (i.e., $B_{11}\ne 0$).
Further, suppose $x_1$ is not critical for some solution to
$(\Phi,A)$. Let $\Phi'$ be the formula obtained by plucking $x_1$ from
$\Phi$. Let $A'$ be the affine space of co-dimension $t-1$ defined by
dropping the first linear equation $\sum_{j=1}^n B_{1j}x_j=b_1$
\emph{after eliminating} $x_1$ from the other linear equations by row
operations. Then $(\Phi',A')$ is satisfiable and \emph{any solution}
$\bar{a'}$ to $(\Phi',A')$ can be extended to a solution $\bar{a}$ of
$(\Phi,A)$.
\end{lemma}

\iffull
\begin{proof}
By assumption, there is a solution $\hat{a}$ to $(\Phi,A)$
  for which $x_1$ is non-critical. Let $\hat{a'}\in\F_2^{n-1}$ be the
  assignment to $x_2,x_3,\ldots,x_n$ obtained from $\hat{a}$ by
  dropping the $x_1$-coordinate. Clearly, $\hat{a'}$ is a solution to
  $(\Phi',A')$. Hence, $(\Phi',A')$ is satisfiable.  Furthermore,
  suppose $\bar{a'}$ is some solution to $(\Phi',A')$.  Then the
  assignment $\bar{a'}$ to the $n-1$ variables $x_2,x_3,\ldots,x_n$
  can be extended by choosing $x_1$ such that the constraint
  $\sum_{j=1}^n B_{1j}x_j=b_1$ is satisfied. The resulting assignment
  $\bar{a}$ satisfies $\Phi$ and all $t$ constraints defining
  $A$.
\end{proof}
\fi

Lemma~\ref{case2to1} describes a pluck/eliminate step applied to the
non-critical variable $x_1$: namely, pluck $x_1$ from $\Phi$ and
eliminate it from the equations describing $A$.

Clearly, for some sequence of $s\le t$ pluck/eliminate steps applied
successively transforms $(\Phi,A)$ to $(\Phi_s,A_s)$ for which Case 1
holds. Since we do not have an efficient test for checking
non-criticality, the algorithm has to do an exhaustive search for the
sequence of $s$ variables to pluck/eliminate. The number of variable
sequences to consider is bounded by $n^t$. However, as we argue in the
next claim, it suffices to consider each unordered subset $U$ of size $s\le t$
variables and apply pluck/eliminate steps to its variables in the
natural order $x_1,\ldots,x_n$. Thus, we can bound the exhaustive
search to $n\choose {\le t}$ subsets of variables. Let $(\Phi_U,A_U)$
be the resulting instance after pluck/eliminate applied to variables
in $U$ in the natural order.

\begin{lemma}\label{case2to1b}
  Let $(\Phi,A)$ be a satisfiable instance of $\subsat{k}$ with
  $\codim(A)=t$. There is a subset $U$ of variables of size at most
  $t$, such that $(\Phi_U,A_U)$ is a satisfiable Case 1 instance of
  $\subsat{k}$.
\end{lemma}

\iffull
\begin{proof}
Suppose $x_{i_1},x_{i_2},\ldots,x_{i_s}$ is a sequence of
  $s\le t$ variables to which the pluck/eliminate steps applied
  results in a satisfiable Case 1 instance $(\Phi_s,A_s)$. Let the $t$
  equations $Bx=b$ define the affine space $A$. The row operations
  applied with the pluck/eliminate steps transforms this system into
  the following equations (also defining $A$):
\begin{equation}
  \ell_j=x_{i_j}, 1\le j\le s \textrm{ and } \ell_j=0, s+1\le j\le t,
\end{equation}
  for affine linear forms $\ell_j, j\in[t]$ in which none of the
  variables $x_{i_1},x_{i_2},\ldots,x_{i_s}$ occur. Moreover, the
  $t-s$ equations $\ell_j, j>s$ define $A_s$, and for every solution
  $\bar{a}$ to $(\Phi_s,A_s)$ all variables occurring in these $t-s$
  equations are critical for $\bar{a}$ w.r.t $\Phi_s$.  

  Now, suppose we apply the pluck/eliminate steps in the natural order
  to the variable subset $U=\{x_{i_1},x_{i_2},\ldots,x_{i_s}\}$
  resulting in $(\Phi_U,A_U)$. Formulas $\Phi_U$ and $\Phi_s$ are
  identical (as both are obtained by plucking variables from $U$).
  The accompanying row operations for the eliminate steps could result
  in a different set of equations (defining $A$): $\ell'_j=x_{i_j},
  1\le j\le s$ and $\ell'_j=0, s+1\le j\le t$. The variables in $U$ do
  not occur in $\ell'_j, j\in[t]$, and the affine space $A_U$ is
  defined by the $t-s$ equations $\ell'_j=0, j>s$. Since any solution
  to these equations uniquely determines the values to the variables
  in $U$, and all equations together define $A$, we can conclude that
  $A_U=A_s$.
\end{proof}
\fi

\smallskip\noindent \textbf{The $O^*({n\choose \le t}\cdot 2^{n-n/k})$ time Algorithm.}

\smallskip\noindent
On input $(\Phi,A)$, the algorithm proceeds as follows:

\medskip \noindent For each subset $U\subset V_\inn$ of size at most $t$ do the following:
\begin{enumerate}
\item Pluck the variables in $U$ from $\Phi$ to obtain $\Phi_U$.
\item For each variable $x_i\in U$ (in any order): pick some equation in
  which $x_i$ occurs; remove $x_i$ from other equations by adding the
  picked equation to it; drop the picked equation from the system.
\item Run the PPZ algorithm on the resulting instance $(\Phi_U,A_U)$
  as if Case 1 were applicable. More precisely, run PPZ on $\Phi_U$
  for $O^*(2^{n-n/k})$ steps; for each solution obtained, if it
  satisfies $A_U$ then output an extension of it to a solution to
  $(\Phi,A)$ and exit,\footnote{From a solution to $(\Phi_U,A_U)$ we
    can reconstruct the solution to $(\Phi,A)$ as the values to
    variables in $U$ are uniquely determined via the linear equations
    from the values to the other variables.} else continue the
  for-loop for the next choice of subset $U$.
 \end{enumerate}

To see the correctness, suppose $(\Phi,A)$ is satisfiable. By
Lemma~\ref{case2to1b}, for some choice of $U$ with $|U| \le t$,
$(\Phi_U,A_U)$ is a Case 1 instance. Hence, the PPZ satisfiability
algorithm will output a solution to $(\Phi_U,A_U)$ in time
$O^*(2^{n-n/k})$ with high probability. This solution can be uniquely
extended to a solution to $(\Phi,A)$ using the linear equations.

We have thus shown the following.
  
 \begin{theorem}
  There is a randomized $O^*({n\choose t}\cdot 2^{n-n/k})$ time algorithm for
  $\subsat{k}$ for subspaces of co-dimension $t$. In particular, for
  $t = o(n)$ we have a randomized $O^*(2^{n(1-1/k+o(1))})$ time
  algorithm.
 \end{theorem}
 
\subsection{An $O^*(2^{n-n/2k+n/2k^2})$ time PPZ-based algorithm for $\subsat{k}$}\label{ppz-algo2}

 Let $(\Phi,A)$ be a $\subsat{k}$ instance. Our objective is a
 randomized algorithm with run time $2^{n-(1-\nu)n/k}$ for as small an
 $\nu$ as possible (ideally, tending to zero).

 To this end, we can first apply Valiant-Vazirani Lemma \cite{VV86} to increase
 the number of constraints (thereby reducing the rank of $A$) and
 getting an instance $(\Phi,A')$ such that $\Phi$ has a unique
 solution in $A'$ with high probability (i.e., inverse polynomial
 probability as guaranteed by Valiant-Vazirani).

 If $\dim(A')\le n-(1-\nu)n/k$ we can brute force search in $A'$ in
 deterministic time $2^{\dim(A')}\le 2^{n-(1-\nu)n/k}$. Thus, we can
 assume that $\dim(A')=n-t$ and $A'$ is the solution space of
 $t<(1-\nu)n/k$ independent affine linear equations.

 Let now $\bar{a}\in\F_2^n$ be the unique solution to the $\subsat{k}$
 instance $(\Phi,A')$.  We partition the variable set into
 $V_\inn\sqcup V_\out$ as before.

 \begin{claim}
   Every variable in $V_\out$ is critical for the satisfying
   assignment $\bar{a}$ of $\Phi$.
\end{claim}

 \claimproof{Suppose $x_i\in V_\out$ is not critical for
   $\bar{a}$. Then $\bar{a}+e_i$ is also a satisfying assignment for
   $\Phi$. Moreover, since $x_i$ does not occur in $V_\inn$,
   $\bar{a}+e_i$ satisfies the linear equations defining $A'$.  Hence
   $\bar{a}+e_i$ is a solution to $(\Phi,A')$ contradicting the
   uniqueness of $\bar{a}$.  }

\medskip\noindent\textbf{The variable plucking algorithm.}
If $\bar{a}$ has more than $(1-\nu)n$ many critical variables ($\nu$
to be fixed in the analysis) then by running the PPZ satisfiability
algorithm \cite{PPZ} for $O^*(2^{n-(1-\nu)n/k})$ iterations we will
find it with high probability.

Otherwise, there are more than $\nu n$ many variables in $V_\inn$ that
are \emph{not critical} for $\Phi$ at $\bar{a}$.
\begin{enumerate}
\item Repeat the following two steps at most $t$ times.  
\item (The plucking step) Randomly pluck a variable $x_i$ from
  $V_\inn$ and drop it from the formula $\Phi$ to obtain its
  \emph{shrinking} $\Phi_1$. Take a linear equation $\ell=b$ in which
  $x_i$ occurs. By row operations eliminate $x_i$ from all other
  linear equations in which $x_i$ occurs and then drop the equation
  $\ell=b$. Let the affine space described by the new set of at most
  $t-1$ linear equations be $A_1$. We claim that $(\Phi_1,A_1)$ also
  has a unique solution $\bar{a}_1$ (obtained from $\bar{a}$ by
  dropping the $i^{th}$ coordinate).
  
\item Let $n_1=n-1$. Run the PPZ algorithm for $2^{n_1-(1-\nu)n_1/k}$
  time on $\Phi_1$. If we do not find the unique solution $\bar{a}_1$
  then repeat the plucking step.
\end{enumerate}

At the end of $t$ successful plucking steps we are left with a $\ksat$
instance $\Phi_t$ with a unique solution (the subspace $A_t$ is
$\F_2^n$) and PPZ will find that solution from which we can compute
$\bar{a}$ by recovering the unique values of the plucked variables
using the linear equations.

\medskip\noindent\textbf{Analysis.} At the $j^{th}$ iteration of the plucking
step, the probability that all $j$ steps pluck off non-critical
variables is at least $\nu^j$.  Thus, the running time of the search
for unique solutions for the $(\Phi_j,A_j)$ over all $t$ steps is
bounded by $\sum_{j=0}^t O^*(\frac{1}{\nu^j}\cdot
2^{n_j-(1-\nu)n_j/k})$.

Letting $\alpha = 2^{1-(1-\nu)/k}$ and noting that $n_j=n-j$ we can
rewrite and bound the above sum as
\begin{eqnarray*}
  O^*(2^{n-(1-\nu)n/k})\cdot \sum_{j=0}^t \frac{1}{\nu^j\cdot \alpha^j}
  & \le & O^*(2^{n-(1-\nu)n/k})\cdot  t\cdot \frac{1}{\nu^t\cdot \alpha^t}\\
  & \le & O^*(2^{n-(1-\nu)n/k})\cdot t\cdot \left(\frac{1}{2\nu}\right)^{(1-\nu)n/k}\cdot 2^{(1-\nu)n/k^2},
\end{eqnarray*}

as the sum $\sum_{j=0}^t \frac{1}{\nu^j\cdot \alpha^j}$ is bounded by
$t \frac{1}{\nu^t\cdot \alpha^t}$ for $\nu\alpha<1$ and $t\le
(1-\nu)n/k$.

The overall running time of the algorithm is, therefore, 
$O^*(2^{n-n/k})\cdot 2^{\nu n/k}\cdot \left(\frac{1}{2\nu}\right)^{(1-\nu)n/k}\cdot 2^{(1-\nu)n/k^2}$,
which is minimized at $\nu=1/2$ as we argue below, and is given by $O^*(2^{n-n/2k+n/2k^2})$.

\iffull
Ignoring the last factor, we need to minimize $2^{\nu n/k}\cdot
\left(\frac{1}{2\nu}\right)^{(1-\nu)n/k}$. In other words, we need to
minimize
\[
2^\nu\cdot\left(\frac{1}{2\nu}\right)^{1-\nu},
\]
Or, equivalently, minimize
\[
\nu\log(4\nu) - \log(2\nu) \textrm{ over }\nu\in[0,0.5].
\]

This is minimized at $\nu=0.5$ and the minimum value is
also $0.5$.
\fi

\begin{remark}[Extension beyond linear-algebraic constraints]
  We note some aspects about the algorithm and explain its adaptation
  to the more general setting of \kcnf\ satisfiability in the
  presence of a global boolean constraint $C(x_1,x_2,\ldots,x_n)$ with
  the property that given a partial assignment to the variables $x_i$
  we can extend the assignment to the remaining variables that
  satisfies the constraint $C$, if such an extension exists. We set
  $\nu =1/2$ and $t=n/2k$. Note that the algorithm need not partition
  the variables into $V_\inn$ and $V_\out$. If there are over $n/2$
  non-critical variables, the algorithm can "obliviously" pluck one
  with probability $1/2$. Oblivious in the sense that it does not need
  to see the constraint $C$. After $t=n/2k$ plucking steps, there are
  at most $n-n/2k$ remaining variables.  We add a final step to the
  algorithm which is a brute-force search over all $2^{n-n/2k}$
  assignments to the remaining variables. For each assignment to these
  that satisfies $\Phi_t$ we can check, in polynomial time, if there
  is an extension to it that satisfies $C$. This search will succeed
  for the unique solution $\bar{a}$. An interesting example for
  constraint $C$ would be $\horn$ formulas. As clause size is
  unrestricted in $\horn$ formulas, notice that neither a direct
  application of the PPZ satisfiability algorithm, nor an application
  of the polynomial equations algorithms would give constant savings
  in the exponent for the runtime bound.
\end{remark} 

More generally, call a Boolean constraint $C(x_1,x_2,\ldots,x_n)$
$T(n)$-easy if there is a $T(n)$ time-bounded algorithm that searches
for a satisfying extension of a given partial assignment to the
variables $x_i$.

\begin{theorem}
 There is a randomized $O^*(2^{n-n/2k+n/2k^2}\cdot T(n))$ time
 algorithm that takes any $\kcnf$ formula and a $T(n)$-easy boolean
 constraint $C(x_1,x_2,\ldots,x_n)$ as input and computes a satisfying
 assignment for the formula and $C$.
\end{theorem}

\begin{corollary}
  There is a randomized $O^*(2^{n-n/2k+n/2k^2})$ time algorithm for
  $\subsat{k}$.
\end{corollary}

\subsection{An $O^*(1.5^r)$ time algorithm for $\subsat{2}$}

\begin{theorem}\label{2subsatalgo}
  Given a $\subsat{2}$ instance $(\Phi,A)$, where $\Phi$ is a $\2cnf$
  formula and $A\subset \F_2^n$ is an $r$-dimensional affine subspace
  given by linear equations, there is a randomized $O^*(1.5^r)$ time
  algorithm to check if $\Phi$ has a satisfying assignment in $A$ and
  if so to compute it.
\end{theorem}

\iffull
\begin{proof}
  Let $X=\{x_1,x_2,\ldots,x_r,\ldots,x_n\}$ be the variable set.
  Without loss of generality, we can assume that $x_1,x_2,\ldots,x_r$
  are independent variables and for $j>r$ we have $x_j=\ell_j$, where
  $\ell_j$ is a linear form in $x_1,x_2,\ldots,x_r$. The literal
  $\bar{x}_j$ is the affine linear form $\ell_j+1$.

  Thus, we can treat the instance $(\Phi,A)$ as a conjunction $\Psi$
  of disjunctions $(\ell \vee \ell')$, where $\ell$ and $\ell'$ are
  affine linear forms in $x_1,x_2,\ldots,x_r$. We can think of this
  satisfiability problem as picking one affine form from each such
  $2$-disjunction $(\ell \vee \ell')$ and setting it to true such that
  the resulting equations are all consistent (i.e. the equations have
  a solution in $\F_2^r$).
  
  We describe below a randomized algorithm that builds a system of
  independent linear equations over $x_1,x_2,\ldots,x_r$ such that any
  satisfying assignment $\bar{a}$ is a solution to this system of
  linear equations with probability at least $(2/3)^r$, and, moreover,
  any solution to this system satisfies $\Psi$. Clearly repeating this
  algorithm $O^*(1.5^r)$ times will find a satisfying assignment to
  the $\subsat{2}$ instance $\Psi$ if one exists.

Here is a description of the algorithm to convert $\Psi$ to a system
of linear equations:
\begin{enumerate}

\item The algorithm runs in stages $i=0,1,\dots$ where in the $i^{th}$
  Stage, it has a system of linear equations $\ell'_j=1, 1\le j\le i$
  for a collection of linearly independent affine forms $\ell'_j$. We
  start off with the empty system at stage $0$.

\item (Stage $i+1$): Take a clause $(\ell \vee \ell')$. If either
  $\ell=1$ or $\ell'=1$ is implied by the equations from stage $i$
  (which can be checked by solving linear equations) then we can
  discard that clause as satisfied and examine the next clause. If
  both $\ell=0$ and $\ell'=0$ are implied by the equations then this
  is a rejecting computation and algorithm outputs ``\textrm{fail}''.

  If $\ell=0$ is implied by the equations and $\ell'$ is independent
  of the $\ell'_j$ then we include the equation $\ell'=1$ and go to
  Stage $i+2$ (if there are any clauses left). Finally, if both $\ell$
  and $\ell'$ are independent of the $\ell'_j$ then we randomly pick
  one of \emph{three} linear forms $\ell, \ell'$ and $\ell + \ell'$,
  include the equation setting it to $1$ and go to Stage $i+2$ (if
  there are any clauses left).

\item Let the final stage be $r'$. Note that $r'\le r$ since the
  equations $\ell'_j=1$ are all independent. At this stage we have no
  clauses left and any solution to the linear equations $\ell'_j=1,
  1\le j\le r'$ satisfies $\Psi$. Output an arbitrary such solution.

\end{enumerate}
We now analyze the success probability of the algorithm.  Suppose
$\bar{a}\in\F_2^r$ is a satisfying assignment for $\Psi$. We claim
that the probability that $\bar{a}$ satisfies the final system of
equations $\ell'_j=1, 1\le j\le r'$ is at least $(2/3)^r$.  We will
prove this by an induction on the stage number $i$: the induction
hypothesis is that $\bar{a}$ satisfies the set of equations at stage
$i$ with probability at least $(2/3)^i$. Clearly, it holds at $i=0$.

For the induction step, suppose after Stage $i$, the assignment
$\bar{a}$ satisfies $\ell'_j=1, 1\le j\le i$. Then notice that in
Stage $i+1$ we either deterministically add the equation $\ell'=1$
which $\bar{a}$ must satisfy since it does not satisfy $\ell=1$
(indeed $\ell$ must evaluate to $0$ at $\bar{a}$), or we randomly pick
one of $\ell, \ell'$ and $\ell+\ell'$. Clearly, $\bar{a}$ must satisfy
exactly two of these three linear forms.  Hence at the end of Stage
$i+1$ the assignment $\bar{a}$ satisfies the system $\ell'_j=1, 1\le
j\le i+1$ with probability at least $(2/3)^{i+1}$. It follows that at
the end of stage $r' \le r$, $\bar{a}$ satisfies the equations with
probability at least $(2/3)^r$.
\end{proof}
\fi

\begin{remark}
  The run time of $O^*(1.5^r)$ that we obtain improves on the
  polynomial equations based algorithms, where for $k=2$ the best run
  time so far is $O^*(1.618^r)$ \cite{D21}.  For $k=3$ a similar randomized branching strategy gives an algorithm
  with run time $O^*((7/4)^r)$. For larger $k$ the run time degrades
  to $O^*((2-1/2^{k-1})^r)$. This runtime bound is obtained similarly
  as for Theorem~\ref{2subsatalgo}: fix a satisfying assignment
  $\bar{a}$ of the $\subsat{k}$ instance. For a clause
  $(\ell_1\vee\ell_2\vee\cdots \vee \ell_k)$ of $k$ linearly
  independent linear forms a random (nonzero) linear combination
  $\sum_{i=1}^k\alpha_i\ell_i$ evaluates to $1$ at $\bar{a}$ with
  probability exactly $\frac{2^{k-1}}{2^k-1}$. 

\end{remark}

\subsection{$\subsat{2}$ in a co-dimension $t$ subspace}\label{subsec:2sat-BGG}
In this section we consider $\subsat{2}$ where we are seeking a
solution in an affine space $A$ such that $\codim(A)=t$.

Given a formula $\Phi$ we will identify a canonical satisfying
assignment $\bar{a}$ for $\Phi$ based on which we will define critical
variables. Since $\twosat$ is in polynomial-time, we can detect
non-critical variables in $\Phi$ w.r.t.\ $\bar{a}$ in polynomial
time. Now the plucking step will try all the possible $n\choose t$
choices of plucking non-critical variables, recalling that a
non-critical variable plucked from a linear constraint defining $A$
allows us to drop that constraint. 

\begin{theorem}\label{2sat}
There is an $O^*({n\choose t})$ time deterministic algorithm for
checking if a $\subsat{2}$ instance $(\Phi,A)$ is satisfiable where
the affine space $A$ has co-dimension $t$.
\end{theorem}

\iffull
\begin{proof}
Let $\Phi$ be a $\2cnf$ formula in variables $x_i,i\in[n]$. 

We first do a standard preprocessing of $\Phi$ by considering its
implication graph on the $2n$ literals $x_i,\overline{x}_i, i\in[n]$,
where for each clause $u\vee u'$, for literals $u$ and $u'$, we have
two directed edges $(\overline{u},u')$ and $\overline{u'},u)$. The
literals that form strongly connected components must all take the
same value in any satisfying assignment and, therefore, can be
replaced by a single variable. This shrinks the implication graph to a
DAG and also reduces the number of variables. Thus, without loss of
generality, we can assume the implication graph of $\Phi$ is a DAG,
and we refer to $\Phi$ as a \emph{reduced} $\twosat$ formula.

\noindent\textit{Computing a canonical satisfying assignment.}~ A
standard linear-time $\twosat$ algorithm computes a canonical
satisfying assignment $\bar{a}$ for $\Phi$ (if satisfiable) by the
following algorithm:
\begin{itemize}
\item[(a)] All literals of outdegree $0$ in the implication DAG are
  assigned true.
\item[(b)] The formula $\Phi$ is simplified after this substitution
  and the new implication DAG computed. If the DAG is non-empty we
  repeat Step(a).
\end{itemize}

The following claim uses the above algorithm to identify non-critical
variables for some satisfying assignment for $\Phi$.

\begin{claim}
 Let $\Phi$ be a $\twosat$ formula with implication DAG $G$. Let
 $u\in\{x_i,\overline{x}_i\}$ be an outdegree $0$ literal in $G$.  If
 $\Phi$ is not satisfiable with $u=0$ then $x_i$ is critical for every
 satisfying assignment of $\Phi$, and if $\Phi$ is satisfiable with
 $u=0$ then $x_i$ is non-critical for every satisfying assignment for
 $\Phi$ that sets $u=0$.
\end{claim}

\claimproof{If there is no satisfying assignment for $\Phi$ with $u=0$
  then clearly $x_i$ is critical for every satisfying
  assignment. Conversely, suppose $\bar{a}$ is a satisfying assignment
  with $u=0$. Then we note that $x_i$ is not critical for $\bar{a}$
  because $\bar{a}+e_i$ is also a satisfying assignment for
  $\Phi$. More precisely, because $u$ has outdegree $0$ in the
  implication graph we can set $u=1$, while retaining the other values
  in $\bar{a}$, and it remains a satisfying assignment.}

More generally, given $\Phi$ we can partition the literals occurring
in its implication DAG $G$ as $S_0\sqcup S_1\sqcup\cdots \sqcup S_w$,
where $S_0$ is the set of outdegree $0$ literals in $G$, $S_1$ is the
set of outdegree $0$ literals in DAG $G_1=G\setminus S_0$, and in
general $S_i$ is the set of outdegree $0$ literals in the DAG
$G_{i+1}=G_i\setminus S_i$. For a variable $x_i$ let $\depth(x_i)$ be
the least index $j$ such that $x_i$ or its complement is in $S_j$.

We observe the following claim which is an easy consequence of the
previous one.

\begin{claim}
  Let $\Phi'$ be the $\twosat$ formula obtained by setting all
  literals in $S_0\sqcup S_i\cdots\sqcup S_{i-1}$ to true. For $u\in
  S_i$, if $\Phi'$ has no satisfying assignment with $u=0$ then $u$ is
  critical for every satisfying assignment for $\Phi$ that sets all
  literals in $S_0\sqcup S_i\cdots\sqcup S_{i-1}$ to true.  If $\Phi'$
  has a satisfying assignment with $u=0$ then $u$ is non-critical for
  every satisfying assignment of $\Phi$ that sets all literals in
  $S_0\sqcup S_i\cdots\sqcup S_{i-1}$ to true.
\end{claim}

We can immediately conclude the following.

\begin{claim}
 If there is a satisfying assignment for $\Phi$ in which all variables
 are critical that has to be the canonical satisfying assignment.
\end{claim}

We describe the basic search procedure used by the algorithm.
\begin{enumerate} 
\item Let $\Phi_0=\Phi$ and $A_0=A$.
\item Repeat the following for steps $s=0$ to $t-1$.
\item Find the canonical satisfying assignment for $\Phi_s$.
\item If it satisfies the linear equations $\ell_i=0, i\in[t-s]$
  defining $A_s$ then output and stop (we can extend it uniquely to
  the $s$ plucked non-critical variables using the linear equations).
\item Else a variable occurring in some $\ell_i$ is non-critical for
  $\Phi_s$ in the solution assignment.
\item Pick a non-critical variable $x_j$ with minimum $\depth(x_j)$
  and pluck it from $\Phi_s$ to get $\Phi_{s+1}$. We take a linear
  equation $\ell_i=0$ where $x_j$ occurs in $\ell_i$, eliminate $x_j$
  from all other equations by row operations using $\ell_i$, and
  finally drop the constraint $\ell_i=0$ to obtain a new affine space
  $A_{s+1}$ . Continue with the repeat step.
\end{enumerate}  
  
Clearly, as long as the canonical satisfying assignment for $\Phi_s$
does not satisfy the system of equations $\ell_i=0$ we can remove a
non-critical variable occurring in one of the $\ell_i$ from $\Phi_s$.

Correctness of the algorithm follows from noting that $(\Phi_s,A_s)$
is satisfiable if and only if $(\Phi_{s+1},A_{s+1})$ is satisfiable, and
if $t$ non-critical variables are plucked then the problem reduces
to a $\twosat$ instances (without any linear constraints).

To complete the overall algorithm, in the basic iteration procedure we
need to cycle through all possible choices of non-critical $x_j$ at
minimum depth $\depth(x_j)$. Since we are going to pluck at most $t$
non-critical variables, this can be done by a brute-force search over
all ${n\choose t}$ subsets of the variables. The running time bound
also follows.
\end{proof}
\fi

\section{Hardness results}\label{sec-hard}

In this section we prove our hardness results for subspace
satisfiability. Since $\sat{k}$ itself is NP-hard for $k\ge 3$, so is
$\subsat{k}$ for $k\ge 3$. So we focus on the case $k=2$.

\subsection{NP-hardness of $\subsat{2}$}

While $\sat{2}$ is polynomial time solvable, the following theorem
shows that $\subsat{2}$ is NP-hard. Note that this follows from
Schaefer's dichotomy theorem for Boolean CSP as the combination of
$\sat{2}$ constraints and linear equations (even with 3 variables per
equation) is not one of the six tractable cases, and thus
NP-hard. Below we give a direct proof based on a simple reduction.

\begin{theorem}\label{thm:np-hard-2subsat}
$\subsat{2}$ is NP-hard.
\end{theorem}

\iffull
\begin{proof}
We show that we can express the NP-hard problem Graph $4$-Colorability
as an instance of $\subsat{2}$, or equivalently $\pafsat{2}$. Indeed,
given a graph $G=(V,E)$, the instance of $\pafsat{2}$ consists of two
Boolean variables $x_{u,1},x_{u,2}$ for each $u \in V$, which will
encode the $2$-bit representation of the $4$ possible colors we can
assign to $u$.  For each edge $e =(u,v) \in E$, we include the
polynomial equation
\begin{equation}
 \label{eq:e-color}
 (x_{u,1} + x_{v,1} + 1) \cdot  (x_{u,2} + x_{v,2} + 1) = 0 \ . 
 \end{equation}
Note that this equation is satisfied iff $x_{u,1} \neq x_{v,1}$ or
$x_{u,2} \neq x_{v,2}$, i.e., when $(x_{u,1},x_{u,2}) \neq
(x_{v,1},x_{v,2})$, which captures the fact the vertices $u$ and $v$
get different colors. The simultaneous satisfiability of the equations
\eqref{eq:e-color} for all $e \in E$ is thus equivalent to $G$ being
$4$-colorable.
\end{proof}
\fi

\subsection{W[1]-hardness of $\subsat{2}$ parameterized by co-dimension}

We now strengthen the hardness result of
Theorem~\ref{thm:np-hard-2subsat} and show that $\subsat{2}$ is
unlikely to even be fixed-parameter tractable when parameterized by
the co-dimension $t$ of the subspace in which we seek a satisfying
assignment to the 2CNF formula.  On the other hand, recall that (as
shown in \cite{BGG-stoc19} and also Section~\ref{subsec:2sat-BGG}),
for fixed co-dimension $t$, $\subsat{2}$ can be solved in polynomial
time. Our $\W[1]$-hardness answers (in the negative) a question posed
in \cite{BGG-stoc19} on whether $\sat{2}$ with a single modular
constraint modulo $M$ is fixed-parameter tractable when parameterized
by $M$ (they gave an algorithm with complexity $n^{O(M)}$).

\begin{theorem}\label{w1-hardness}
Consider the $\subsat{2}$ where the input subspace within which one
has to satisfy the $\sat{2}$ formula has co-dimension
$t$. Parameterized by $t$, $\subsat{2}$ is $\W[1]$-hard.
\end{theorem}

\iffull
\begin{proof}
We give a reduction from the problem \multclique. The input to
\multclique\ consists of a graph $G$, an integer $t$, and a partition
$(V_1,V_2,\dots,V_t)$ of the vertices of $G$, and the task is to
decide if there is a $t$-clique in $G$ containing exactly one vertex
from each part $V_i$. The parameter associated with the problem is
$t$. The problem \multclique\ parameterized by $t$ is known to be
W[1]-hard \cite[Lemma 1]{FHRV}.

The variables in the $\subsat{2}$ instance correspond to the vertices
of the graph. Let us denote these variables by $x_v$ for $v \in V :=
V_1 \cup V_2 \cup \cdots \cup V_k$. The 2CNF clauses in the instance
will be the following:
\begin{itemize}
    \item For all $i\in \{1,2,\dots,t\}$ and $v \neq v' \in V_i$, the clause $(\neg x_v \vee \neg x_{v'})$. These clauses  ensure that at most one $x_v$ can be set to $1$ in each part.
    \item If $(u,v)$ is not an edge in the graph with $G$, the clause $(\neg x_u \vee \neg x_v)$. These clauses ensure that the set $\{ u \mid x_u = 1\}$ must induce a clique in $G$.
\end{itemize}
Note that this instance of $\sat{2}$ is trivial to satisfy by setting all variables to $0$. The affine space $A$ we will use to make this an instance of $\subsat{2}$ is defined by the following equations:
\begin{equation}
\label{eq:2sat-aff}
\sum_{u \in V_i} x_u =1 \quad \text{for } i=1,2,\dots,t \ .
\end{equation}
We stress that the above equations are over $\F_2$, and thus stipulate
that there are an odd number of variables set to $1$ in each part. But
together with the 2CNF clauses which ensure that at most one variable in
each part can be set to $1$, it follows that satisfying assignments of this
$\subsat{2}$ instance are in one-one correspondence with $t$-cliques
of $G$ that include exactly one vertex from each $V_i$. The proof is
now complete by noting that the co-dimension of the affine space $A$
defined by \eqref{eq:2sat-aff} equals $t$. Parameterizing
$\multclique$ by the clique size is thus equivalent to parameterizing
the constructed $\subsat{2}$ instance by the co-dimension.
\end{proof}
\fi

\subsection{Approximability of Max-$\subsat{2}$}

Given the hardness of deciding exact satisfiability of $\subsat{2}$
instance, we now turn to approximate satisfiability. In the
$\maxsubsat$ problem, the goal is to satisfy the maximum number of
2SAT clauses with an assignment that belongs to the input affine space
$A$. Thus, the affine constraints are treated as hard constraints. We
allow clauses of width 1. If unary clauses are disallowed in the 2CNF
formula, and each clause involves exactly two distinct variables, we
call the problem $\maxesubsat$.

\subsubsection{Easy approximation algorithms}
We can assume that no variable is forced to $0$ or $1$ by the affine space $A$, since if that happens we can just set and remove that variable and work on the reduced instance. 
If we pick a random assignment from $A$, it will satisfy at least $1/2$ of the clauses of the 2CNF formula in expectation, and in fact at least an expected fraction $3/4$ of the clauses when each clause involves two distinct variables. The algorithms are easily derandomized. 
For satisfiable instances of \maxsubsat, one can find a $3/4$ approximate solution, as one can eliminate all the unary clauses, and add those conditions to the subspace inside which we want to find an assignment to the 2CNF formula. 
So we get the following trivial algorithmic guarantees.

\begin{observation}
\label{obs:algo}
In polynomial time, one can get a factor $1/2$ approximate solution to instances of \maxsubsat, a factor $3/4$ approximate solution to instances of \maxesubsat, and a factor $3/4$ approximate solution to satisfiable instances of \maxsubsat.
\end{observation}
\noindent 
We will now show that all the above guarantees are best possible, with matching NP-hardness results.

\subsubsection{Tight inapproximability via simple reductions}
For the hardness results and rest of the section, it is convenient to work with the $\paf$ formulation of $\subssat$. The Max-LIN2 problem, of maximizing the number of satisfied equations in a system of affine equations mod 2, trivially reduces to \maxpafsat\ (with each equation being degree 1 instead of degree 2). By H\aa stad's seminal tight inapproximability for Max-LIN2, we have the following.

\begin{observation}
\label{obs:hardness-1}
For any $\epsilon > 0$, \maxpafsat\ (and thus \maxsubsat) is NP-hard to approximate within a factor of $(1/2+\epsilon)$, and this holds for almost satisfiable instances that admit an assignment satisfying a fraction $(1-\epsilon)$ of equations.
\end{observation}

We also get a tight hardness (matching Observation~\ref{obs:algo}) for the \maxesubsat\ or equivalently when each polynomial equation is the product of exactly two (linearly independent) affine forms.

\begin{lemma}
\label{lem:lin-to-paf}
For any  $\epsilon > 0$, \maxepafsat\ is NP-hard to approximate within a factor of $(3/4+\epsilon)$, and this holds for almost satisfiable instances that admit an assignment satisfying a fraction $(1-\epsilon)$ of equations.
\end{lemma}

\iffull
\begin{proof}
This follows from a simple reduction from Max-LIN2. Suppose we are given a system of affine equations $A_1=0,A_2=0,\cdots,A_m=0$, where the $A_i$'s are distinct affine forms in Boolean variables $x_1,x_2,\dots,x_n$. We produce a system of ${m \choose 2}$ quadratic equations $A_i \cdot A_j = 0$ for $1 \le i < j \le m$ in the same variables $x_1,x_2,\dots,x_n$. If an assignment to the $x_i$'s violates $r$ affine constraints $A_j=0$, then the same assignment violates ${r \choose 2}$ of the quadratic constraints. When $r = \epsilon m$, the fraction of violated quadratic constraints is $\approx \epsilon^2$, and when $r = 1/2-\epsilon$,   the fraction of violated quadratic constraints is $\approx 3/4 - O(\epsilon)$. The claimed hardness now follows from H\aa stad's inapproximability result for Max-LIN2.
\end{proof}
\fi

\subsubsection{Inapproximability for satisfiable instances}
\label{sec:satisfiable-hardness}
The above inpproximability results do not apply to \emph{satisfiable} instances of $\subsat{2}$. They are obtained by reductions from linear equations whose exact satisfiability can be easily checked. We now prove that approximating \maxsubsat\ doesn't get easier on satisfiable instances.

\begin{theorem}
\label{thm:tight-hardness}
For every $\epsilon > 0$, 
it is NP-hard to approximately solve satisfiable instance of \maxesubsat\ within a factor of $3/4+\epsilon$. That is, it is NP-hard to find, given as input a satisfiable instance of \subsat{2}, an assignment satisfying a fraction $3/4+\epsilon$ of the 2SAT constraints.
\end{theorem}
\iffull
\begin{proof}
Consider the arity 3 Boolean CSP which is defined by the predicate $\oxr : \{0,1\}^3 \to \{0,1\}$, defined by 
\[ \oxr(x_1,x_2,x_3) = x_1 \vee (x_2 \oplus x_3) \]
applied to literals. En route his celebrated tight inapproximability for satisfiable Max-3SAT, H\aa stad proved that the CSP defined by $\oxr$ (and with negations allowed on variables) is NP-hard to approximate within a factor of $(3/4+\epsilon)$ even on satisfiable instances, for arbitrary $\epsilon > 0$. (Note that independent random choices of the bits $x_1,x_2,x_3$ makes $\oxr(x_1,x_2,x_3)=1$ with probability $3/4$, so the hardness factor of $3/4$ is tight.) Now the constraint $\oxr(x_1,x_2,x_3)=1$ is equivalent to the equation
\[ (x_1 + 1) (x_2 + x_3 +1) =0 \]
stipulating that a product of two affine forms vanishes. Thus the CSP defined by $\oxr$ can be equivalently expressed as a $\subsat{2}$ instance, and the claimed inapproximability of $\maxesubsat$ on satisfiable instances follows.
\end{proof}
\fi

\section{System of polynomial equations over binary field: effect of reducibility}\label{sec-algos2}

We now examine a special case of the problem of solving a system of
polynomial equations over $\F_2$ studied in
\cite{LPTWY17,BKW19,D21}. For motivating background, we recall
according to the strong exponential time hypothesis (SETH) that
$\SAT$, that is $n$-variable CNF satisfiability of unrestricted clause
width, cannot be essentially solved faster than $2^n$ time. However,
Schuler~\cite{Sch05} and Calabro et al~\cite{CIP06} have shown the
special case that sparse instances of $\SAT$ (with $c\cdot n$ clauses)
can be solved in $O^*(2^{n(1-\alpha)})$ time, where $\alpha$ is a
constant depending on the clause density $c$. It is natural to ask if
there is an analogous result for $\Subsat$ (satisfiability of
conjunctions of unbounded disjunctions of affine linear forms). In
this section we show a more general algorithmic result in the setting
of systems of polynomial equations over $\F_2$.

Let $P_i\in\F_2[x_1,x_2,\ldots,x_n], 1\le i\le m$ be polynomials over
the field $\F_2$ as input instance to the $\poleq$ problem. The
problem is denoted $\polyeq{k}$ when the degrees are bounded by $k$
which generalizes $\subsat{k}$ as already explained in the
introduction.

The unrestricted degree case is significantly different, because we
can easily combine the $m$ equations into a single equation as
follows.  Define
\[
P = 1 + \prod_{i=1}^m (1+P_i).
\]
Clearly, the system $P_i=0, 1\le i\le m$ has a solution iff $P=0$ has
a solution. 

Thus, assuming SETH, there is no algorithm essentially faster than
$2^n$ for solving $P=0$.

\begin{remark}
  There is also the question of how the polynomials $P_i$ are given as
  part of the input. If $\deg P_i\le k$ for all $P_i$ then we can in
  polynomial-time compute their sparse representation as a linear
  combination of the $n^k$ many monomials of degree at most $k$.
  However, in the above reduction of combining the $P_i$ into a single
  polynomial, $P$ is a small arithmetic formula.  In fact, for the
  case of $\poleq$ we consider, where the instance is a system of
  equations $P_i=0, 1\le i\le m$ such that $m=O(n)$ and each $P_i$ has
  constant degree irreducible factors, we can assume that the $P_i$
  are given as arithmetic circuits.
\end{remark}

We now show that $\poleq$ instances $P_i=0, 1\le i\le m$ can be solved
faster than $2^n$ if $m$ is linear in $n$ and the irreducible factors
of each $P_i$ are of constant degree. This can be seen as a
``polynomial equations'' analogue of Schuler's $\SAT$ algorithm for
spare instances with unrestricted clause width \cite{Sch05,CIP06}. We note that a different degree reduction method, based on a rank argument, is used in \cite[Section 4]{LPTWY17} to solve systems of polynomial equations $p_i=0$, where each $p_i$ is given by a sum of product of affine linear forms.

\begin{theorem}\label{thm:schuler-type}
  Let $P_i=0, 1\le i\le c \cdot n$, for a constant $c>0$, be an
  instance of $\poleq$, such that the degree of each irreducible
  factor of each $P_i$ is bounded by a constant $b$. There is a
  randomized algorithm for $\poleq$ that runs in time
  $2^{n(1-\alpha)}$ for such instances, where $\alpha>0$ is a constant
  that depends on $c$ and $b$.
\end{theorem}
\iffull
\begin{proof}
  We can factorize each polynomial $P_i$ into its irreducible factors in randomized polynomial time using Kaltofen's algorithm \cite{K89}. Let
\[
P_i=\prod_{j=1}^{r_i} Q_{ij}
\]
be this factorization for each $i$. Define polynomials
$R_{ij}=1+Q_{ij}$ for each $i$ and $j$, and note that $\deg R_{ij}\le
b$. For $a_{ijs}\in \F_2$ picked independently and uniformly at random
define polynomials
\[
\tilde{R}_{is} = \sum_{j=1}^{r_i}a_{ijs}R_{ij}, 1\le s \le \log m + 2.
\]
Finally, we define the polynomials 
\[
\tilde{R}_i= \prod_{s=1}^{(\beta+1)\log c}(1+ \tilde{R}_{is}), 1\le i \le m,
\]
where $\beta>0$ is a constant to be fixed later in the analysis.

Notice that $\deg \tilde{R}_i \le b \cdot (\beta+1)\log c$ for each $i$.

\begin{claim}
  If $P_i=0, 1\le i\le m$ is unsatisfiable then $\tilde{R}_i=0, 1\le i\le m$ is also unsatisfiable.
\end{claim}

To see this, suppose $P_i(\bar{a})=1$ at assignment
$\bar{a}\in\F_2^r$. Then $Q_{ij}(\bar{a})=1$ for each $j$ which
implies each $R_{ij}(\bar{a})=0$ for each $j$. It follows that
$\tilde{R}_{is}=0$ for all $s$ and hence $\tilde{R}_i=1$.

On the other hand, we have:

\begin{claim}
  If $\bar{a}\in\F_2^n$ is a solution to the system of equations
  $P_i=0, 1\le i\le m$ then with probability at least $e^{-n/c^\beta}$
  $\bar{a}$ is a solution to the sytem of equations $\tilde{R}_i=0,
  1\le i\le m$.
\end{claim}

The probability that $\bar{a}$ is a solution to the single equation
$\tilde{R}_i=0$ is given by $1-\frac{1}{c^{\beta+1}}$. Since the
events are independent, the probability that $\bar{a}$ is a solution
to the system $\tilde{R}_i, 1\le i\le m$ is given by
\begin{eqnarray*}
(1-\frac{1}{c^{\beta+1}})^m & = & (1-\frac{1}{c^{\beta+1}})^{cn}\\
                         & \approx & e^{-n/c^\beta}.
\end{eqnarray*}

Now the system of equations $\tilde{R}_i, 1\le i\le m$ is an instance
of $\polyeq{k}$, where $k=b(\beta+1)\log c$ is a constant. Applying
one of the algorithms \cite{LPTWY17,BKW19,D21} yields an
$O^*(2^{n(1-1/2k)})$ algorithm with success probability
$e^{-n/c^\beta}$. We can boost the success probability to a constant
with an overall run time of $O^*(2^{n(1-1/2k)}\cdot e^{n/c^\beta})$,
which can be optimized by choosing $\beta$ appropriately.
\end{proof}
\fi

\section*{Acknowledgment}
We thank anonymous reviewers for useful comments and pointers to the literature.

\bibliographystyle{plainurl}


\end{document}